\documentclass[showpacs,nofootinbib,aps]{revtex4}

\usepackage{amssymb}
\usepackage{amsmath}
\usepackage{graphicx}  

\usepackage{color}

\usepackage[normalem]{ulem}

\usepackage[dvipsnames]{xcolor}

\usepackage{setspace}

\newcommand{\mathsym}[1]{{}}
\newcommand{\unicode}[1]{{}}

\usepackage{tablefootnote}

\usepackage{braket}

\usepackage{physics}

\usepackage{soul}

\begin{document}

\title{Measurement-induced entanglement entropy of gravitational wave detections}

\author{Preston Jones}
\email{Preston.Jones1@erau.edu}
\author{Quentin G. Bailey} 
\author{Andri Gretarsson}
\author{Edward Poon}

\affiliation{Embry Riddle Aeronautical University, Prescott, AZ 86301}


\date{\today}

\begin{abstract}
Research on the projective measurement of gravitons increasingly supports Dyson’s conclusions that the detection of single gravitons is not physically possible. It is therefore prudent to consider alternative signatures of non-classicality in gravitational wave detections to determine if gravity is quantized. Coincident multiple detector operations make it possible to consider the bipartite measurement-induced entanglement, in the detection process, as a signature of non-classicality. By developing a model of measurement-induced entanglement, based on a fixed number of gravitons for the bipartite system, we demonstrate that the entanglement entropy is on the order of a few percent of the mean number of gravitons interacting with the detectors. The bipartite measurement-induced entanglement is part of the detection process, which avoids the challenges associated with developing signatures of production-induced entanglement, due to the extremely low gravitational wave detector efficiencies. The calculation of normalized measurement-induced entanglement entropy demonstrates the potential of developing physically meaningful signatures of non-classicality based on bipartite detections of gravitational radiation. This result is in stark contrast to the discouraging calculations based on single-point detections.
\end{abstract}


\maketitle

\let\endtitlepage\relax

\section{Introduction}\label{Introduction}

Dyson has made a compelling case \cite{Dyson13} that the detection of single gravitons is not physically possible. This represents a serious challenge for efforts to determine the quantum nature of gravity and has inspired a great deal of ongoing research \cite{Rothman06,Lieu18,Parikh21PRD,Lake23,Carney24,Palessandro24}, which supports Dyson's conclusions. The physical restrictions on the detection of single gravitons excludes one but only one measure of non-classically. Broadly speaking, bosonic non-classicality is associated with a relatively small number of processes e.g., single particle detection, particle bunching, quadrature squeezing, sub-Poissonian correlations \cite{Giovannini11,Giovannini17,Giovannini19,Kanno19,Kanno20}, intensity interferometry \cite{Jones23}, weak measurement \cite{White16,Tilloy19,Athira21}, and state entanglement. The focus of this paper is on the entanglement of identical gravitons in the processes of two-point detections: specifically, the measurement-induced entanglement entropy \cite{Tichy11,Schroeder17,Franco16,Franco18,Heaney21} for gravitational wave detections with detectors that are physically separated but operating concurrently \cite{Jones23,Parikh23} and not sufficiently separated to exclude overlap \cite{Peres95,Shi03} of the state functions.

The problem of identifying the quantum character of gravitational radiation differs from that of quantum optics where the existence of the photon has been well established. The projective measurement of photon state functions is an unequivocal demonstration of non-classicality that makes the question of alternative signatures of non-classicality a matter of an appropriate detector schema and sensitivity. Since the comparable projective measurement of graviton quantum states is very likely to be physically impossible, other methods to demonstrate non-classicality (and by inference the existence of the graviton) are required. The problem of theory development is complicated by this absence of any certain measure demonstrating that gravitational radiation is not completely classical. While we would expect that the theory of gravito-optics \cite{Jones23} would closely follow similar theories in quantum optics we cannot know a priori that the theory represents physical reality. In order to make meaningful progress in developing physically realizable signatures of non-classicality for gravitational radiation we relate theory from quantum optics to bipartite gravitational wave detections and measurement-induced entanglement entropy as an alternative to projective measurement of single graviton state functions.

Our principle goal is the calculation of measurement-induced entanglement entropy as a quantitative measure of non-classicality.  The magnitude of the calculated entropy for bipartite detections provides an encouraging measure of potential signatures of non-classicality in gravitational wave detections. Importantly, this does not rely on the (physically impossible) projective measurement of single gravitons in the detection process. Alternative measures of non-classicality, based on bipartite detection, would also be associated with state function entanglement and we would expect the magnitude of any such comparable measures to be similar. While the magnitude of measurement-induced entanglement entropy does illustrate the potential of two-point detection processes for the demonstration of non-classicality in gravitational wave detections \cite{Vedral98,Dodonov02} entropy is generally difficult to measure directly. However, the bipartite entanglement would also be expected to manifest in more readily measurable phenomena e.g., in increased coincidence rates and many other detector responses associated with the bipartite detection process as discussed in Section \ref{Detection}.

The calculation of measurement-induced entanglement entropy for gravitational wave detections follows other recent works on entanglement entropy as a measure of gravitational wave non-classicality. Related to astrophysical gravitational wave detections and contemporary or near-future detector processes, ``linearized gravity" was considered as a potential mechanism for gravity-induced entanglement \cite{Nandi24} in a quantum oscillator. In the context of cosmological production, primordial gravitational waves were shown to induce entanglement and the associated measure of entanglement entropy in graviton to photon conversion processes \cite{Kanno23}. The production processes for primordial gravitational waves have also been shown to develop entanglement entropy for relic gravitons \cite{Giovannini24}. Both mechanisms of primordial graviton entanglement could be detectable by future space-based detectors. Supporting this recent research, the present work provides a quantitative measure of bipartite entanglement of graviton state functions, produced during the detection process, which would not be subject to decoherence and other challenges associated with exceedingly low detector efficiencies \cite{Jones23,Carney24} as discussed in the following section. 

While the entropy of a system is generally difficult to measure it is at the same time an excellent tool for the characterization of the state of a system. Our principle goal is to demonstrate the viability of detecting non-classicality for gravitational radiation by calculating the measurement-induced entanglement entropy for a bipartite gravitational wave detection. We find that the entropy calculation for contemporary gravitational wave detectors could be as great as a few percent of the detector response relative to the number of effective graviton interactions. As strain sensitivity improves for future detectors the normalized measurement-induced entanglement entropy, as a potential signature of non-classicality, will also improve.

\section{Mean number of detector gravitons} \label{efficiency}

In terms of quantum measurements, many contemporary and future gravitational wave detectors are based on interactions of an optical cavity with a gravitational wave signal. Based on the exceedingly small graviton-photon scattering cross-section \cite{Skobelev75,Rothman06} it is not surprising that projective measurement of the gravitational radiation \cite{Lieu18,Carney24,Palessandro24} does not contribute to the detection of gravitational waves for optical cavities. As previously discussed the calculations of the interactions of the gravitational wave detectors with the signal show that the gravitons interact weakly with the detectors. The following semi-classical model of the gravitational wave detection demonstrates that the extremely weak interaction of gravitons with e.g. photons in the optical cavity is inconsistent with the representation of the gravitational wave detection as the sum of individual graviton interactions with the detector. This would represent a non-projective measurement of gravitational radiation, weakly interacting with the detector, and is characteristic of quantum weak measurements \cite{Carney24}. 

Our present purpose is to determine if measurement-induced entanglement can meaningfully contribute to the detection processes for gravitational waves. Toward this end we take a semi-classical approach to modeling the interactions of the detectors with the gravitational radiation. That is, we assume that the interaction of quantized gravitational radiation with the detector is non-projective and treat the single detector response as classical.

Following Lieu \cite{Lieu18,Chen18,Jones23} the average detector response energy for a single arm can be characterized in terms of the deviation from equilibrium $\xi = h L e^{- \omega t}$, as $ E =  \frac{1}{2} M \dot{\xi}^2 = \frac{1}{8} M h^2 L^2 \omega^2$  \cite{Lieu18,Jones23}. Taking the energy per graviton to be  $ E_{g}  =  \hbar \omega $ and the effective detector mean graviton number per cycle for two detector arms to be $E_{tot}=2E$, the mean single detector graviton number is

 \begin{equation} 
 \bar{n} = \frac{ E_{tot} }{E_{g}} =\frac{  \frac{1}{4} M h^2 L^2 \omega^2 }{ \hbar \omega } =   \frac{1}{4 \hbar}   M L^2 \omega  h^2.
 \label{gravitonNumber}
\end{equation}

\noindent We assume mirror masses $M = 50 ~ \rm{kg} $, detector arm lengths $L = 4,000 ~ \rm{m} $, and e.g., for $\omega = 10^3 ~  \rm{\frac{rad}{s}}$ the mean detector graviton number is $ \bar{n} = \left(  \frac{1}{4 ~ 10^{-34} ~ \rm{Js}} 50 ~ \rm{kg} \left( 4,000 ~ \rm{m} \right)^2 10^3 ~  \rm{\frac{rad}{s}} \right) h^2  = \left(2 \times 10^{45}   \right) h^2 $. As an example when $ h = 10^{-21} $ the mean detector graviton number would be $\bar{n} = 2 \times 10^{3}$. In order to estimate the entanglement in bipartite detections we will approximate the relation between strain amplitude and effective number using this estimate. While the proceeding ``back of the envelope'' calculation is Newtonian the general relativistic characterization of the detector response differs by a gauge condition \cite{Chen18}, justifying the our use of the Newtonian calculation.

The effective detector mean graviton number can also characterize detector efficiencies. The average energy flux per cycle \cite{Jones23}, $E_{tot} \frac{\omega}{2 \pi L^2}$, is used to estimate the detector efficiency,

  \begin{equation} 
 \eta =\frac{  \frac{1}{8 \pi} M h^2 \omega^2 }{ F_g } \sim \frac{G}{c^3} M \omega,
 \label{efficeincy}
\end{equation}

\noindent where $F_g =  \frac{c^3 h^2 \omega^2}{16 \pi G} $ is the gravitational wave flux. Assuming typical values for contemporary detectors the efficiencies are on the order of $\eta \sim 10^{-31}$. These extremely low theoretical efficiencies present a serious challenge for detection of production-induced entanglement \cite{Carney24} for both contemporary and future detectors. The contrasting theoretical limit for measurement-induced entanglement is calculated in Section \ref{Mentanglement}, as the ratio of the entanglement entropy and detector response, and is potentially statistically meaningful for both contemporary and future detectors.

\section{Measurement-induced entanglement}\label{entanglement}

\begin{figure}[htp!]
 \centering
\includegraphics[width=75mm]{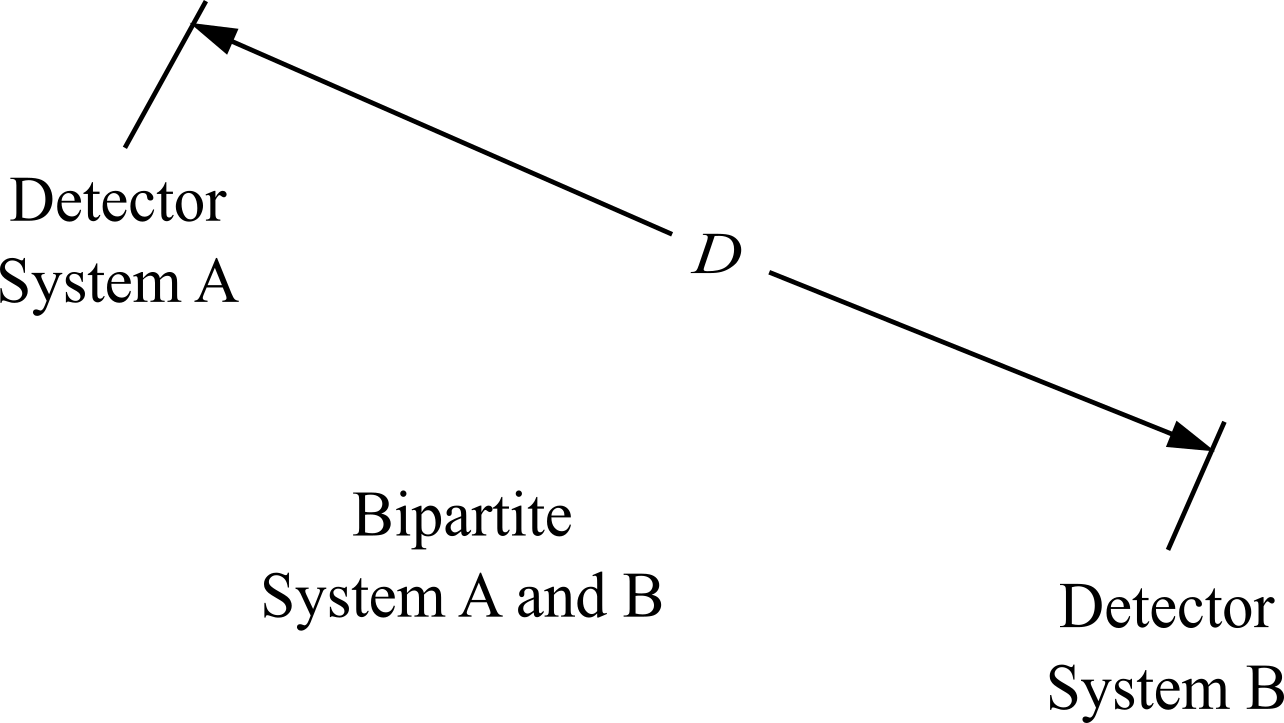}
 \caption{Measurement-induced entanglement requires overlap of the graviton state functions \cite{Peres95},  over the separation distance $D$ between detectors A and B. Any Earth based bipartite detector system would have a physical separation on the order of the graviton state function wavelength e. g., $ D = c / 100 ~\rm{Hz} \sim 10^6 ~\rm{m}$ and satisfy this requirement. The detectors would also be required to operate concurrently.}\label{BipartitDectectos}
\end{figure}

In order to estimate the relative magnitude, as shown in Fig. \ref{EntropyVsNumber}, of potential non-classicality we construct a semi-classical model for a measurement-induced entangled boson state \cite{Franco16,Franco18} with subsystems $A$ and $B$ and fixed total number $n$. Consistent with the weak detector response to quanta of the gravitational radiation the detector will be considered classically. The bipartite detector states are entangled by the measurement of indistinguishable gravitons in a superposition of detection at subsystems $A$ or $B$ as illustrated in Fig. \ref{BipartitDectectos}.




Gravitational quanta are spin 2 and the number states can be characterized in terms of creation and annihilation operators. Following Kanno and Soda \cite{Kanno20} we would define a number state for the gravitational radiation with cross and plus polarizations in terms of two sets of the graviton creation and annihilation operators, $ \hat{b}_\oplus , ~ \hat{b}_\oplus^{\dag},$ and $ \hat{b}_\otimes, \hat{b}_\otimes^{\dag}$. We can simplify our model by considering a single polarization of the gravitational radiation and associated number states. With this assumption the number states of the subsystems are formally the same as optical states, $\hat{b} \ket{n}  = \sqrt{n}\,\ket{n-1} $ and $\hat{b}^{\dag} \ket{n} = \sqrt{n+1}\,\ket{n+1}$. Consideration of a single polarization is appropriate here where the focus is on measurement-induced entanglement for two-point detections. Gravitational wave detections including both cross and plus polarizations could also produce identical particle entanglement. This could be interesting in future studies for both single point detections \cite{Nandi24} as well as two-point detections.

Assuming a single polarization of the gravitational radiation the state function \cite{Dalton17} for bipartite entanglement with fixed total number is,

\begin{equation}
\ket{\varphi_{n}} = \sum_{k=0}^{n} c \left(n, k \right) \ket{k}_{A}  \ket{n - k}_{B} ~,
\label{genericfixed}
\end{equation}

\noindent where $n$ is the total number of particles for the bipartite system. We will also assume that detector subsystems $A$ and $B$ are symmetric with mean detector number $ \bar{n}_A = \bar{n}_B = \bar{n} $ and total fixed particle number $ n= n_A + n_B = 2 \bar{n}$, where $\bar{n}$ is the mean detector graviton number \eqref{gravitonNumber}. The bipartite state of subsystems $A$ and $B$ is a pure state since it is a state function or vector in a Hilbert space. For a pure state the density operator is $\hat{\rho} = \ket{\varphi_{n}}  \bra{\varphi_{n}}$, or in an orthonormal basis \cite{Gerry05},

\begin{equation}
\hat{\rho} =   \sum_{j=0}^{n}  \sum_{k=0}^{n}   c \left(n, k \right) c^{*} \left(n, j \right) \ket{k}_{A}  \ket{n - k}_{B} \bra{n - j}_{B}  \bra{j}_{A} ~.
\label{bipartiteDMgen}
\end{equation}

\noindent To determine the entanglement entropy for the subsystems \cite{Dalton17} we need to calculate the partial trace for $A$ or $B$ from \eqref{bipartiteDMgen} where $\hat{\rho}_A =\hat{\rho}_B$. The resulting entanglement entropy represents a measure of the system entanglement that is locally accessible \cite{Navarrete15} to separate observers for subsystem $A$ or $B$. The expansion of the partial trace with respect to $B$ can be developed as $\hat{\rho}_A = {\rm tr}_B \left( \hat{\rho} \right)$ and more explicitly,






\begin{equation}
\hat{\rho}_A = \sum_{l=0}^{n} \bra{n-l}_B  ~ \, \hat{\rho} ~ \, \ket{n-l}_B ~,
\label{bipartiteDMpartialA3}
\end{equation}






\noindent which simplifies to

\begin{equation}
\hat{\rho}_A = \sum_{l=0}^{n}  c \left(n, l \right) c^{*} \left(n, l \right) \ket{l}_{A}  \bra{l}_{A}  ~.
\label{bipartiteDMpartialA5}
\end{equation}

\noindent For a bipartite system the entanglement entropy can be calculated from the spectral decomposition of \eqref{bipartiteDMpartialA5}, $\lambda \left( k \right) = c \left(n, k \right) c^{*} \left(n, k \right) $ for subsystem $A$ or $B$ e.g.,

\begin{equation}
S_A  = - {\rm tr} \hat{\rho}_A \left( {\rm ln} ~ \hat{\rho}_A \right) = -  \sum_{k=0}^{n}  \lambda \left( k \right) {\rm ln} \left(  \lambda \left( k \right) \right) ~.
\label{EEntropySpectral}
\end{equation}

\noindent  If there is no measurement-induced entanglement, in the bipartite detection process, the entanglement entropy $S_A=0$ \cite{Chong21}.

\section{Entanglement entropy}\label{Mentanglement}

In order to develop a model state function for the fixed number measurement-induced entanglement for a bipartite system we will need to determine the occupation amplitudes $ c \left(n, k \right)$, which are in general complex. Consider the example of the occupation amplitudes for a coherent state,

\begin{equation}
\ket{\alpha} = e^{- \frac{1}{2} \abs{\alpha}^2 } \sum_{k=0}^{ \infty} \frac{\alpha^k}{\sqrt{k !}} \ket{k } ~.
\label{coherent}
\end{equation}

\noindent We start our model development, for measurement-induced entanglement, from the probability amplitudes of a coherent state \eqref{coherent} by taking $\alpha^k \rightarrow \bar{n}^{\frac{k}{2}} e^{i \phi_k}$ and $ \abs{\alpha}^2 \rightarrow \bar{n} $  \cite{Dalton17} in \eqref{genericfixed},

\begin{equation}
\ket{\varphi_{n}}_p = \mathcal{N}_p \left( n , \bar{n} \right) e^{- \frac{1}{2}  \bar{n}}   \sum_{k=0}^{n}   \frac{ {\bar{n}^{\frac{k}{2}} e^{i \phi_k}} }{\sqrt{k !}} \ket{k}_{A}   \ket{n - k}_{B} ~,
\label{bipartitefixed}
\end{equation}

\noindent where $\mathcal{N}_p \left( n, \bar{n} \right)$ is a normalization. The square of the normalization has a closed form solution in terms of gamma functions, $\mathcal{N}^2_p \left( n, \bar{n} \right) =   \frac{\Gamma\left(n + 1 , \right) }{\Gamma\left(n + 1, \bar{n} \right) }  $.  The probability amplitudes in \eqref{bipartitefixed} are Poissonian and will be approximately symmetric for large $n$. Collecting terms the elements of the spectral decomposition are $\lambda_p \left( k \right) =  \frac{\Gamma\left(n + 1 , \right) e^{- \bar{n}}  }{\Gamma\left(n + 1, \bar{n} \right) } \frac{ {\bar{n}^{k} } }{k !} ~$.

\begin{figure}[htp!]
 \centering
\includegraphics[width=85mm]{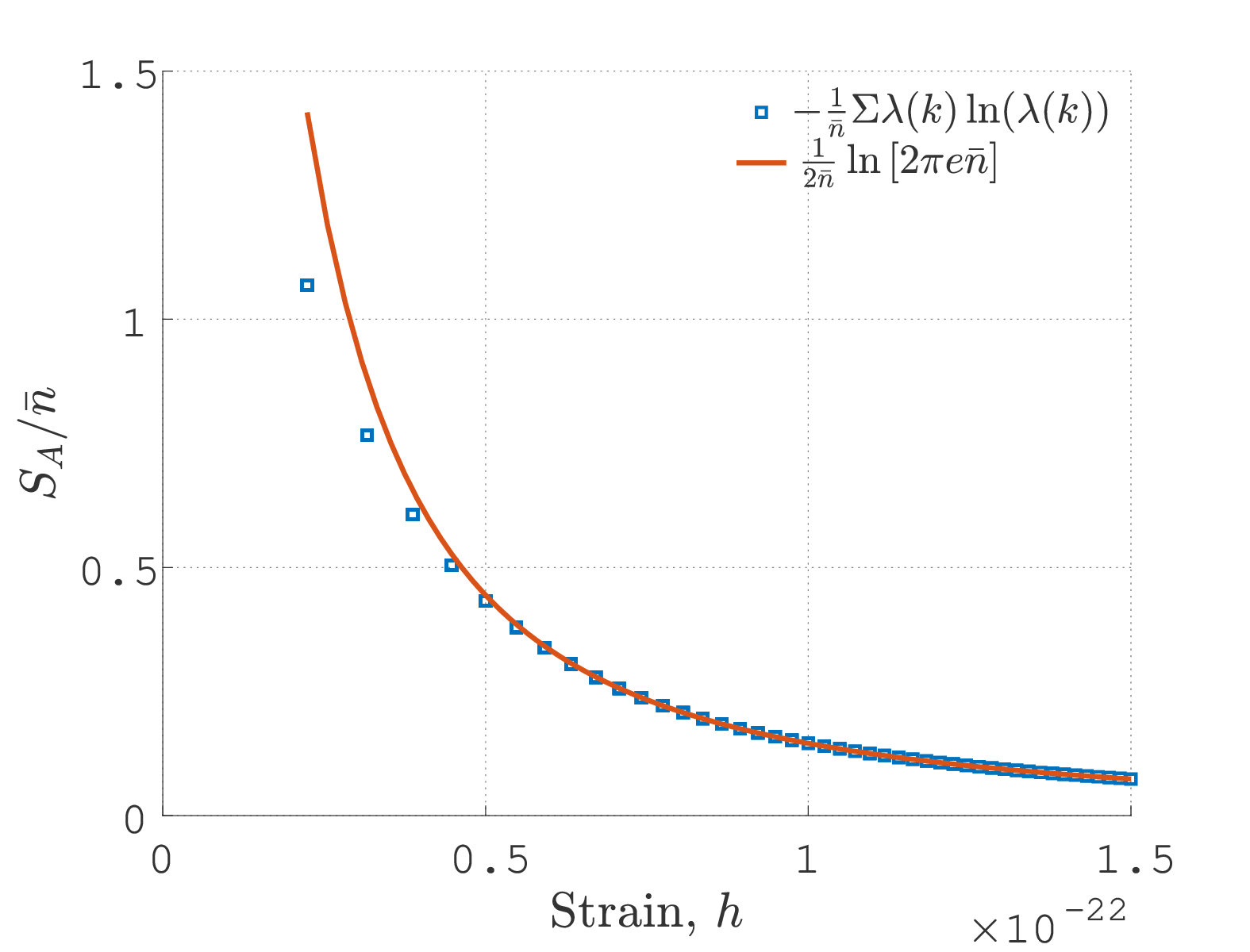}
 \caption{The measurement-induced entanglement entropy, $S_A$ from \eqref{EEntropySpectral} and \eqref{SpecDecomG}, is plotted as a fraction of the detected effective mean graviton number $ \bar{n} = \left(  2 \times 10^{45} \right) h^2 $ in terms of the gravitational wave strain amplitude $h$. The range shown is within the sensitivity of contemporary and near future detectors. The curve fit \eqref{S3} is a good estimate for values of $h > 5 \times 10^{-23}$.}\label{EntropyVsNumber}
\end{figure}

The Poisson spectral decomposition does provide a model for entangled biparite states but the states are not symmetric. To construct a probability distribution that is symmetric for systems $A$ and $B$ assume Gaussian amplitudes in \eqref{genericfixed}, $c \left( n, k \right) = \mathcal{N}_g \left( n , \bar{n} \right) e^{ - \frac{ \left(k - \bar{n} \right)^2}{4 \sigma^2}} e^{i \phi_k} $,

\begin{equation}
\ket{\varphi_{n}}_g = \mathcal{N}_g \left( n , \bar{n} \right)  \sum_{k=0}^{n}  e^{ - \frac{ \left(k - \bar{n} \right)^2}{4 \sigma^2}} e^{i \phi_k}  \ket{k}_{A}  \ket{n - k}_{B} ~,
\label{Gbipartitefixed}
\end{equation}

\noindent with normalization $ \mathcal{N}^2_g = \left( \sum_{j=0}^{n}   e^{ - \frac{ \left(j - \bar{n} \right)^2}{2  \sigma^2}} \right)^{-1}$ and elements of the spectral decomposition, 

\begin{equation}
\lambda_g \left( k \right) =  \left( \sum_{j=0}^{n}   e^{ - \frac{ \left(j - \bar{n} \right)^2}{2  \sigma^2}} \right)^{-1} e^{ - \frac{ \left(k - \bar{n} \right)^2}{2  \sigma^2}}  ~,
\label{SpecDecomG}
\end{equation}

\noindent where we take $\sigma^2 = \bar{n}$ consistent with the Poisson states \eqref{bipartitefixed} for large $\bar{n}$. The asymmetry in the Poissonian probability amplitudes is unphysical and we will adopt the Gaussian probability amplitudes which are symmetric. We plot the measurement-induced entanglement entropy for fixed total particle number from \eqref{EEntropySpectral} and \eqref{SpecDecomG} in Fig. \ref{EntropyVsNumber} in the nominal range of gravitational wave detector sensitivities.

The plot in Fig. \ref{EntropyVsNumber} includes a curve fit for larger values of $\bar{n}$, where the curve fit is determined by the asymptote. The asymptote of the measurement-induced entanglement, in Fig. \ref{EntropyVsNumber}, can be developed by setting $m = \bar{n}$, $t_k = e^{-(k-m)^2/2m}$, and $T = \sum_{k=0}^{2m} t_k$.  Then $\lambda_g(k) = t_k/T$ and the entanglement entropy \eqref{EEntropySpectral}

\begin{equation}\label{Entropy}
S_A = \ln T - \frac{1}{T} \sum_{k=0}^{2m} t_k \ln t_k.
\end{equation}

\noindent Now we estimate some sums to establish asymptotics.  By viewing $\sum_{j=1}^m e^{-j^2/2m}$ as a Riemann sum, we have


\begin{equation}\label{S0}
\sqrt{2m} \int\limits_{\frac{1}{\sqrt{2m}}}^{\frac{m+1}{\sqrt{2m}}} e^{-u^2} du = \int\limits_1^{m+1} e^{-\frac{x^2}{2m}} dx \leq \sum_{j=1}^m e^{-\frac{j^2}{2m}} \\ \leq \int\limits_0^{m} e^{-\frac{x^2}{2m}} dx = \sqrt{2m} \int\limits_0^{\sqrt{\frac{m}{2}}} e^{-u^2} du,
\end{equation}

\noindent and consequently, by the squeeze theorem,

\begin{equation}\label{S1}
\lim_{m \to \infty} \frac{1}{\sqrt{2m}}  \sum_{j=1}^m e^{-j^2/2m}  = \int\limits_0^{\infty} e^{-u^2} du = \frac{ \sqrt{\pi}}{ ~ 2}.
\end{equation}

\noindent Noting that $f(x) = x^2 e^{-x^2/2m}$ has a maximum value of $2m/e$ at $x=\sqrt{2m}$, and viewing $\sum_{j=1}^m j^2 e^{-j^2/2m}$ as a Riemann sum, we have


\begin{equation}
\int\limits_0^{m+1} f(x)  dx - \frac {2m}{e} \leq \int\limits_0^{m+1} f(x) \, dx - \int\limits_{\lfloor \sqrt{2m} \rfloor}^{\lceil \sqrt{2m} \rceil} f(x)  dx \\ \leq \sum_{j=1}^m j^2 e^{-\frac{j^2}{2m}} \leq \int\limits_1^{\lfloor \sqrt{2m} \rfloor} f(x)  dx + \int\limits_{\lceil \sqrt{2m} \rceil}^m f(x) dx + \frac{4m}{e},
\end{equation}








\noindent where $\lfloor x \rfloor$ is the greatest integer $p \leq x$, and $\lceil x \rceil$ is the least integer $q \geq x$.  Consequently

\begin{equation}\label{S2}
\lim\limits_{m \to \infty} \frac{1}{(2m)^{3/2}}  \sum_{j=1}^m j^2 e^{-\frac{j^2}{2m}}  = \int\limits_0^{\infty} u^2 e^{-u^2} du = \frac{ \sqrt{\pi}}{ ~ 4}.
\end{equation}


\noindent Now $T = \sum_{j=-m}^m e^{-j^2/2m} = 1 + 2 \sum_{j=1}^m e^{-j^2/2m} \sim \sqrt{2\pi m}$ by \eqref{S1}, while $-\sum_{k=0}^{2m} t_k \ln t_k = 2 \sum_{j=1}^m \frac{j^2}{2m} e^{-j^2/2m} \sim \sqrt{m\pi/2}$ by \eqref{S2}. Substituting these asymptotics into \eqref{Entropy} gives

\begin{equation}\label{S3}
S_A \sim \frac{1}{2} \ln (2 \pi e m).
\end{equation}

\noindent Recalling that $m = \bar{n}$ and after normalization, we plot this asymptote along with the normalized measurement-induced entanglement entropy in Fig. \ref{EntropyVsNumber}.

\section{Detection of measurement-induced entanglement}\label{Detection}

Entanglement entropy is an excellent and widely used measure of the state of a system and, in the case of measurement-induced entanglement, an excellent measure of non-classicality for a system. However, in general, it is difficult to measure the entropy of any system and we describe several possibilities for the measurement of entanglement associated with the entropy calculations in Section \ref{Mentanglement}. The magnitude of the entanglement entropy developed in Section \ref{Mentanglement} suggest that the entangled state of the system should in principle be distinguishable from noise, which justifies the consideration of established techniques from quantum-optics as potential protocols in gravito-optics for identification of signatures of non-classicality.

One potentially practical method for detection of the entangled state of a bipartite system is Hanbury Brown and Twiss (HBT) interferometry \cite{Ragy13,Jones23} and in particular measurements of intensity correlations. It is possible that two or more detectors operating as a HBT interferometer would exhibit discernible increases in detection coincidence rates \cite{Jeltes07} when detectors are operating coincidentally. In quantum-optics it is possible to directly measure system entropy or the related quantum purity in applications of quantum information theory. This suggest that research in the development of comparable experimental techniques in gravito-optics could provide comparable measurements of system entanglement for gravitational wave detections. There are many other potential ways of detecting the signatures of non-classicality associated with the measurement-induced entanglement calculated in Section \ref{Mentanglement} and several possibilities are presented in Table \ref{table:Emeasurement}.

Most of the proposed measurement procedures associated with entanglement entropy in TABLE \ref{table:Emeasurement} would be possible with the development of an appropriate schema for the analysis of recorded observations from gravitational wave detections without modification of the detectors. Notable exceptions are the possibility of experiments to directly measure gravity induced entanglement and new protocols introducing atom interferometers \cite{Krisnanda20,Asenbaum20,Carney21,Carney21PRX}. The introduction of atom interferometers to the gravitational wave detection processes would require the development of new experiments operating in conjunction with the gravitational wave detectors. An intriguing feature of these experiments would be that the coupling to entanglement induced across detectors is over distances of thousands of kilometers between independent systems $A$ and $B$. This large physical separation would insure that any measured correlations are completely independent of standard model fields and associated noise.

\section{Discussion - contemporary and future detectors}\label{discussion}

As demonstrated in Section \ref{efficiency}, theoretical graviton detector efficiencies for contemporary and future detectors are and will remain extremely low. This is due to the relatively small number of gravitons interacting during the detection process compared to the graviton flux of the signal. The relatively small mean number of gravitons interacting with the detectors justifies the semi-classical model of the detection process in Section \ref{Mentanglement}, where the signal is treated non-classically and the detector response classically. For this relatively small number of gravitons, the bipartite measurement-induced entanglement is found to be on the order of a percent of the detector response as shown in Fig. \ref{EntropyVsNumber} and demonstrates the potential importance of measurement-induced entanglement in the bipartite detector system and in the development of signatures of non-classicality. It is noteworthy that the detectors are not responding to individual gravitons through projective measurement, which suggest that the consideration of weak measurement theory  \cite{White16,Tilloy19,Athira21} could provided a better understanding of the potential non-classicality in the bipartite detection of gravitational waves. The model developed in Section \ref{Mentanglement} demonstrates that improved sensitivity of future detectors will increase the normalized measurement-induced entanglement entropy as a fraction of the detector response. This should in turn make it easier to discern the increased coincidence rates for multiple detectors.

\begin{widetext}

\noindent \begin{table}[ht]
\begin{tabular}{l|c|c}
 & Description & References \\
\hline

\begin{minipage}[t]{0.15\columnwidth}%
\vspace{5pt} 
 \begin{singlespace*}
 \flushleft \noindent Entanglement entropy  %
\end{singlespace*}
\vspace{10pt}
\end{minipage} & \begin{minipage}[t]{0.65\columnwidth}%
\vspace{5pt} 
 \begin{singlespace*}
\flushleft  Techniques for direct measurement of entanglement entropy and quantum purity are possible in quantum-optics and would be an unambiguous demonstration of non-classicality if a comparable schema could be developed in gravito-optics.%
\end{singlespace*}
\vspace{10pt}
\end{minipage} & \cite{Eisenberg04,Nakazato12,Islam15,Zhang24} \\
\hline

\begin{minipage}[t]{0.15\columnwidth}%
\vspace{5pt} 
 \begin{singlespace*}
\flushleft Squeezed states  %
\end{singlespace*}
\vspace{10pt}
\end{minipage} &  \begin{minipage}[t]{0.65\columnwidth}%
\vspace{5pt} 
 \begin{singlespace*}
\flushleft Measurements of squeezed states for the gravitational wave signals would be an unambiguous demonstration of non-classicality. %
\end{singlespace*}
\vspace{10pt}
\end{minipage}  & \cite{Giovannini19,Kanno19,Kanno23,Giovannini24} \\
\hline

\begin{minipage}[t]{0.15\columnwidth}%
\vspace{5pt} 
 \begin{singlespace*}
\flushleft Intensity correlations   %
\end{singlespace*}
\vspace{10pt}
\end{minipage} &  \begin{minipage}[t]{0.65\columnwidth}%
\vspace{5pt} 
 \begin{singlespace*}
\flushleft Hanbury Brown and Twiss intensity interferometry and multi-detector intensity correlations provide a measure that can be associated with increased coincidence rates due to bosonic entanglement. %
\end{singlespace*}
\vspace{10pt}
\end{minipage}  & \cite{Malvimat14,Jones23} \\
\hline

\begin{minipage}[t]{0.15\columnwidth}%
\vspace{5pt} 
 \begin{singlespace*}
\flushleft Residual noise %
\end{singlespace*}
\vspace{10pt}
\end{minipage} &  \begin{minipage}[t]{0.65\columnwidth}%
\vspace{5pt} 
 \begin{singlespace*}
\flushleft Residual noise in the waveform extraction associated with increased coincidence rates for gravitational wave detector operations, that are not remote, could be associated with entanglement. %
\end{singlespace*}
\vspace{10pt}
\end{minipage}  & \cite{Parikh21PRD,Parikh23} \\
\hline

\begin{minipage}[t]{0.15\columnwidth}%
\vspace{5pt} 
 \begin{singlespace*}
\flushleft Cross and plus polarizations %
\end{singlespace*}
\vspace{10pt}
\end{minipage} &  \begin{minipage}[t]{0.65\columnwidth}%
\vspace{5pt} 
 \begin{singlespace*}
\flushleft Multimode entanglement could be possible with cross and plus polarizations for the systems as well as the subsystems $A$ and $B$. %
\end{singlespace*}
\vspace{10pt}
\end{minipage}  &  \cite{Kanno20,Nandi24} \\
\hline

\begin{minipage}[t]{0.15\columnwidth}%
\vspace{5pt} 
 \begin{singlespace*}
\flushleft Gravity-induced entanglement  %
\end{singlespace*}
\vspace{10pt}
\end{minipage} & \begin{minipage}[t]{0.65\columnwidth}%
\vspace{5pt} 
 \begin{singlespace*}
\flushleft Entanglement of the interferometer mirrors, for both the system and subsystems, due to gravity-induced entanglement, could provide a meaningful signature of non-classicality.  %
\end{singlespace*}
\vspace{10pt}
\end{minipage} & \cite{Matsumura20,Carney21PRX,Gunnink23,Carney24,Liu24,Krisnanda20} \\
\hline

\begin{minipage}[t]{0.15\columnwidth}%
\vspace{5pt} 
 \begin{singlespace*}
\flushleft Atom interferometry %
\end{singlespace*}
\vspace{10pt}
\end{minipage} &  \begin{minipage}[t]{0.65\columnwidth}%
\vspace{5pt} 
 \begin{singlespace*}
\flushleft Atom interferometers, coupled to various gravitational systems, have proven to be useful in tests of classical gravity and similar coupling could be possible for systems of gravitational wave detections. %
\end{singlespace*}
\vspace{10pt}
\end{minipage}  &  \cite{Krisnanda20,Asenbaum20,Carney21,Carney21PRX} \\
\hline
\end{tabular}
\caption{Potential experimental schemata for demonstration of signatures of non-classicality that would be associated with the measurement-induced entanglement entropy described in Section \ref{Mentanglement}.}
\label{table:Emeasurement}
\end{table}

\end{widetext}

While we have considered fixed number measurement-induced entanglement, another theoretical possibility is multi-mode entanglement \cite{Nandi24} associated with cross and plus gravitational wave polarizations as described in Table \ref{table:Emeasurement}. At present gravitational wave detectors are not capable of bipartite detections of cross and plus polarization states \cite{Callister17}, but this could change for future detectors. Multi-mode entanglement would be complementary to fixed number entanglement we consider here and could provide an additional tool for identification of signatures of non-classicality for future detectors.

Importantly, for future detectors, due to the bipartite nature of the entanglement, the measurement-induced entanglement does not introduce local non-classicality restrictions on the potential sensitivity of single-point detectors. However, the measurement-induced entanglement entropy for near future detectors presented in Fig. \ref{EntropyVsNumber} could be significant in terms of the relatively small effective number of gravitons interacting with the detectors. Future detectors with sensitivities on the order of $h=10^{-23}$ would be expected to exhibit clearly discernible signatures of bipartite entanglement associated with the significant normalized measurement-induced entanglement entropy.

\section{Conclusions}\label{conclusions}

With the growing support for Dyson's conclusions that detection of single graviton events are physically impossible, it is prudent to consider alternative signatures of non-classicality in detections of gravitational radiation. Single-point detector demonstrations of signatures of non-classicality, in the incident gravitational wave signal, are further complicated by exceeding low detector efficiencies. Two important steps forward were made here in the theory development for signatures of non-classicality that provide a clear alternative to physically impossible projective measurements in single-point processes. First, is recognizing the potential of measurement-induced entanglement for identical particles and overlapping state functions, where two gravitational wave detectors are operating coincidentally, i.e. not remote. Any pair of Earth-based gravitational wave detectors will satisfy these restrictions as shown in Fig. \ref{BipartitDectectos}. Second, is the demonstration that the magnitude of the normalized measurement-induced entanglement entropy for bipartite detections is on the order of up to a few percent and should be discernible from the noise for appropriate measurement schemata associated with the entanglement.

By calculating the measurement-induced entanglement entropy for bipartite detections we demonstrated that evidence of the existence of the graviton, that are associated with the detection process for gravitational radiation, does not require unphysical detector sensitivities. The entanglement of the bipartite system during the detection process is found to produce potentially detectable signatures of non-classicality based on the magnitude of the normalized measurement-induced entanglement entropy, as shown in Fig. \ref{EntropyVsNumber}. This suggests that practical methods for the determination of the measurement-induced entanglement of the bipartite gravitational wave detections should be possible and warrants further consideration.

The success of the semi-classical model presented here, treating the detectors as classical and the signal as quantized, suggest that future consideration of the bipartite detections as weak measurements could greatly improve our understanding of potential signatures of non-classicality in gravitational wave detections.


\end{document}